\documentclass[a4paper,11pt]{article}
\usepackage{pos}
\usepackage{braket}
\usepackage{subcaption}

\newcommand{\RFT}{Hansen:2014eka,Hansen:2015zga}
\newcommand{\ThreeBodyPapers}[0]{%
Detmold:2008gh,
Beane:2007qr,
Briceno:2012rv,
Polejaeva:2012ut,
Hansen:2014eka,
Hansen:2015zga,
Briceno:2017tce,
Hammer:2017uqm,
Konig:2017krd,
Hammer:2017kms,
Mai:2017bge,
Briceno:2018mlh,
Briceno:2018aml,
Blanton:2019igq,
Pang:2019dfe,
Jackura:2019bmu,
Briceno:2019muc,
Romero-Lopez:2019qrt,
Hansen:2020zhy,
Blanton:2020gha,
Blanton:2020jnm,
Pang:2020pkl,
Romero-Lopez:2020rdq,
Blanton:2020gmf,
Muller:2020vtt,
Blanton:2021mih,
Muller:2021uur,
Blanton:2021eyf,
Jackura:2022gib,
Hansen:2024ffk,
Yan:2024gwp,
Schaaf:2024qer}
\newcommand{\KtoM}{%
Hansen:2015zga,
Jackura:2020bsk,
Dawid:2023kxu,
Jackura:2023qtp,
Briceno:2024ehy}
\newcommand{\UnphysicalStates}{Briceno:2018mlh}

\title{Implementing the RFT finite-volume formalism for three pions across all non-maximal isospins}
\ShortTitle{Implementing the RFT formalism for non-maximal isospin $\pi\pi\pi$}

\author*{Athari Alotaibi}
\author{Maxwell T. Hansen}

\affiliation{Higgs Centre of Theoretical Physics, School of Physics and Astronomy,\\
The University of Edinburgh, Edinburgh, UK}

\emailAdd{A.Alotaibi@sms.ed.ac.uk}
\emailAdd{maxwell.hansen@ed.ac.uk}

\abstract{We present a numerical investigation of the relativistic-field-theoretic (RFT) formalism, used to predict the discrete energy spectrum of three pions in a finite volume. Applying the generalization of ref.~\cite{Hansen:2020zhy}, we extract results for all non-maximal isospin values ($I_{\pi\pi\pi} = 2,1,$ and $0$), for different total momenta $\boldsymbol{P}$, and for various irreducible representations of the finite-volume symmetry group. We restrict attention to the unphysical scenario in which the three-particle interactions are set to zero. This set-up thus serves as a baseline for future lattice QCD calculations that will aim to extract such three-body interactions.}

\FullConference{The 41st International Symposium on Lattice Field Theory (LATTICE2024)\\
28 July - 3 August 2024\\
Liverpool, UK\\}

\begin{document}
\maketitle

\section{Motivation and Overview}

In recent decades, lattice QCD has emerged as a powerful method for reliably predicting the properties of quantum chromodynamics (QCD). The approach is based on numerical evaluation of the QCD path integral on a discretized Euclidean spacetime with a finite temporal extent, typically the longest direction, of length $T$ and three finite spatial extents, typically equal and each of length $L$. One application of the method is to numerically estimate finite-volume Euclidean two-point correlation functions, and from these to calculate finite-volume energies with a given set of fixed quantum numbers. Following the pioneering work of L\"uscher \cite{Luscher:1986n1,Luscher:1986n2}, these can then be related to physical scattering observables.

L\"{u}scher originally developed the formalism for two identical spin-zero particles with zero total momentum. Over the years, the method has been generalized in the two-particle sector to include non-zero total momentum in the finite-volume frame \cite{Rummukainen:1995vs,Kim:2005gf,Christ:2005gi}, as well as multiple channels of non-identical particles, including those with intrinsic spin~\cite{Lage:2009zv,Bernard:2010fp,Fu:2011,Doering:2011,Hansen:2012tf,Briceno:2012yi,Gockeler:2012yj,Briceno:2014oea}. The approach has been successfully applied to numerous two-particle systems, e.g.~to evaluate the masses and couplings of resonances. Recent reviews can be found in refs.~\cite{Briceno:2017max,Mai:2021lwb,Hanlon:2024fjd}.

However, the L\"uscher method is only valid over a fixed range of center-of-mass frame (CMF) energy. In the case of a single scalar field with two-to-three couplings, the upper limit is given by $E^\star < 3m$, where $E^\star$ is the total energy in the CMF and $m$ is the particle mass. In the case where there is no even-odd coupling, this extends to $E^\star < 4m$. It is therefore of great interest to extend the method to describe systems with more than two particles. A natural step for this extension is the inclusion of three-particle channels, relevant for the omega resonance ($\omega(782) \to \pi\pi\pi$) and the Roper resonance ($N(1440) \to \pi\pi N$), as well as to weak decays such as $B\to \pi\pi K$. Extensive progress has been made in this goal. See, for example, refs.~\cite{\ThreeBodyPapers}.

In this talk, we focus on one of these approaches, the so called relativistic-field-theory (RFT) formalism, first presented in refs.~\cite{\RFT}. Our aim here is to provide an efficient implementation and some first results of the RFT method, for three-pion systems with non-maximal isospin, using the formalism derived in ref.~\cite{Hansen:2020zhy}. The spectrum for the maximal isospin, $I_{\pi\pi\pi} = 3$, has been studied more extensively, e.g. in refs.~\cite{Mai:2018djl,Horz:2019rrn,Blanton:2019vdk,Mai:2019fba,Fischer:2020jzp,Hansen:2020otl}. We review some details of this formalism in the following section, before discussing our implementation and results.

\section{Summary of the RFT quantization condition}

The RFT quantization condition, valid up to exponentially suppressed terms scaling as $e^{-m_\pi L}$, where $m_\pi$ is the pion mass (or more generally the lightest mass in the system) is given by
\begin{equation}
\text{det} \Big [1+\mathcal{K}_{{\rm df},3}(E^\star) F_3(E, \boldsymbol{P},L \vert \mathcal{K}_2) \Big ] =0 \,,
\label{eq:QC}
\end{equation}
where $E$ and $\boldsymbol{P}$ are the total energy and momentum of the three particles, related to the CMF energy by $E^\star = \sqrt{E^2 - \boldsymbol P^2}$. The time extent, $T$, is assumed to be large enough that any thermal effects are small and can be neglected. The finite spatial volume is implemented by imposing periodic boundary conditions on the fields, which leads to discretization of all momenta, including the total momentum: $\boldsymbol{P}= (2 \pi/L)\boldsymbol{n}$ where $\boldsymbol{n} \in \mathbb{Z}^3$ is a three-vector of integers. Throughout this work we use the shorthand $\boldsymbol P = [n_xn_yn_z]$ to represent specific momenta, e.g.~$\boldsymbol P = [001]$ indicates the momentum $\boldsymbol P = (2\pi/L)\hat{\boldsymbol z}$.

The quantization condition relates the finite-volume effects encoded in $F_3(E, \boldsymbol P, L \vert \mathcal K_2)$ to the three-particle K-matrix, which describes the short-distance interaction and can be related to the physical scattering amplitude $\mathcal{M}_3$ via known integral equations.%
\footnote{More precisely $F_3$ carries some infinite-volume information since it includes the two-particle subprocess in its definition, via $\mathcal K_2$. Unlike the two-particle quantization condition, the separation between the finite-volume kinematics and the infinite-volume dynamics is not given as a simple product of two factors.}
More details on the relation between $\mathcal{K}_{{\rm df},3}$ and $\mathcal{M}_3$ can be found in refs.~\cite{\KtoM}.

For given values of the two- and three-particle K-matrices, $\mathcal K_2$ and $\mathcal K_{{\rm df}, 3}$, the finite-volume spectrum is predicted as the discrete set of roots of eq.~\eqref{eq:QC} in $E$, for each value of $L$ and $\boldsymbol P$. The solutions, called $E_n(L)$, can be found numerically, making use of the definition~\cite{Hansen:2014eka,Hansen:2020zhy}
\begin{equation}
\label{F3_general}
F_3(E, \boldsymbol P, L \vert \mathcal K_2) \equiv \frac{1}{3}\frac{F}{2\omega L^3} - \frac{F}{2\omega L^3} \frac{1}{1+\mathcal{K}_2[F+ G] } \mathcal{K}_2 F \,,
\end{equation}
where $F$, $G$ and $\omega$ are known kinematic functions.

\begin{figure}[t]
\centering
\includegraphics[width=0.8\textwidth]{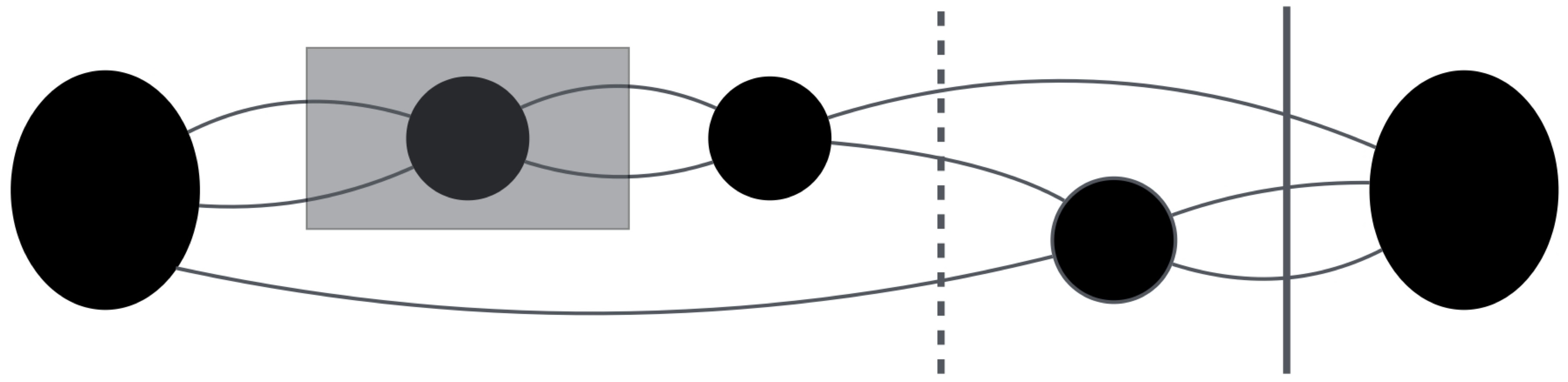}
\caption{As is described in detail in various references, e.g.~refs.~\cite{Hansen:2014eka,Hansen:2019nir,Hansen:2020zhy}, the quantization condition is derived via a skeleton expansion in which all three-particle states are exposed. In the example shown, one can identify the scattering pair as those lines connected to a circle with four legs and spectator as the third particle. This changes at various locations in the diagram. The vertical lines correspond to the finite-volume cuts $F$ and $G$ for \textit{solid} and \textit{dashed} lines, respectively. The shaded square represents a contribution to $\mathcal{K}_2$.}
\label{fig:3p_config}
\end{figure}

The quantities $\mathcal K_{{\rm df}, 3}$, $F_3$, $F$, $G$, $\mathcal K_2$ and $\omega$ are all matrices with indices $\boldsymbol k', \ell', m', \boldsymbol k, \ell, m$. This index space results from considering the three-particle system in terms of a scattering pair (with angular momentum $\ell$) and a spectator particle (with momentum $\boldsymbol k$). If the spectator carries spatial momentum $\boldsymbol{k}$, then the scattering pair has CMF energy
\begin{equation}
E^\star_{2,k} = \sqrt{ (E-\omega_k)^2 - (\boldsymbol{P}-\boldsymbol{k})^2} \,.
\end{equation}
In this frame, the individual particles in the pair have back-to-back spatial momenta with four-momenta given by $(E^\star_{2,k}/2, \boldsymbol{a}^\star)$ and $(E^\star_{2,k}/2, -\boldsymbol{a}^\star)$, where we use the superscript $\star$ throughout to denote quantities that are boosted to a two- or three-particle CMF. The subscript $k$ denotes the choice of spectator particle, which can vary within a given Feynman diagram, as shown in figure~\ref{fig:3p_config}. This figure also illustrates the origin of the three key building blocks within $F_3$: $F$, $G$ and $\mathcal K_2$.

\section{Incorporating non-maximal isospin}

In the original work of refs.~\cite{Hansen:2014eka,Hansen:2015zga}, the quantization condition was implemented for maximal isospin $I_{\pi\pi\pi} = 3$, which is equivalent to the case of three identical $\pi^+$. The extension to non-maximal isospins was derived in ref.~\cite{Hansen:2020zhy}. This was achieved by first considering a basis of pions with definite individual pion flavors ($\pi^0$, $\pi^+$, $\pi^-$) and a total charge of zero for the three-pion state. This leads to two possibilities: $\vert \pi^+ \pi^- \pi^0 \rangle$ and $\vert \pi^0 \pi^0 \pi^0 \rangle$. Assigning definite momenta to the individual pions (e.g.~$p_1$, $p_2$ and $p_3$) then leads to seven possibilities, six from permutations of $\vert \pi^+ \pi^- \pi^0 \rangle$ and one from $\vert \pi^0 \pi^0 \pi^0 \rangle$.

In ref.~\cite{Hansen:2020zhy}, the quantization condition was derived by studying all diagrams like those in figure~\ref{fig:3p_config}, but with these definite pion charges included everywhere. This leads to an extension of eq.~\eqref{eq:QC} in which, in addition to the $\boldsymbol k, \ell , m $ index space, all objects also carry a flavor index $f \in \{ 1, \cdots , 7 \}$ tracking the seven possible neutral three-pion states.

In a final step, the matrix entering the determinant of the quantization condition can be block-diagonalized into the four possible values of the total isospin: $I_{\pi\pi\pi} = 3, 2,1$ and $0$, yielding four independent quantization conditions. We refer the reader to ref.~\cite{Hansen:2020zhy} for full details of the derivation.

\section{Implementation}

The practical implementation of the quantization condition is available as the open-source code `{\sf ampyL}' (\emph{am-pie-ell}) on {\sf github.com}~\cite{ampyL}. Some key aspects of the implementation are as follows:
\begin{itemize}
\item[(i)] Projection of the quantization condition to the irreducible representations (irreps) of the appropriate finite-volume symmetry group. The group is defined as all elements of the octahedral group (including parity) that leave the total momentum $\boldsymbol{P}$ invariant. Table~\ref{tab:irrep_list} lists some basic properties of the symmetry group for the three total momenta considered here.

\begingroup
\setlength{\tabcolsep}{12pt}
\renewcommand{\arraystretch}{1.2}
\begin{table}[b]
\centering
\begin{tabular}{|c|c|c|}
\hline \hline
$\boldsymbol{P}$ & $\mathcal{G}\ (\vert \mathcal G\vert)$ & irrep (dimension) \\
\hline\hline
$[000]$ & $O_h^D(48)$ & $A^\pm_1(1), A^\pm_2(1), E^\pm(2), T^\pm_1(3)$ and $T_2^\pm(3)$ \\
$[001]$ & Dic$_4(8)$ & $A_1(1), A_2(1), B_1(1),B_2(1)$ and $E_2(2)$ \\
$[011]$ & Dic$_2(4)$ & $A_1(1), A_2(1), B_1(1)$ and $B_2(1)$ \\
\hline \hline
\end{tabular}
\caption{Finite-volume symmetry groups for three values of the total momentum, $\boldsymbol P$, written in units of $(2\pi/L)$. $\mathcal{G}$ and $\vert \mathcal G\vert$ denote the name of the group and the number of elements, respectively.}
\label{tab:irrep_list}
\end{table}
\endgroup

\item[(ii)] Prediction of the non-interacting finite-volume energies and multiplicities for a specific set of three-pion quantum numbers. These are given by the sum of the individual pion energies, $E_{\pi}(\boldsymbol p_1) + E_{\pi}(\boldsymbol p_2) + E_{\pi}(\boldsymbol p_3)$, where $E_{\pi}(p_i) = \sqrt{m_\pi^2 + \boldsymbol p_i^2}$.

\item[(iii)] Extraction of all relevant building blocks (e.g.~$F$, $G$, $\mathcal K_2$) for a definite total isospin channel. Following ref.~\cite{Hansen:2020zhy}, the three-pion system decomposes as
\begin{equation}
1\otimes 1 \otimes 1 =
1_{\sigma\pi}\oplus (0\oplus 1\oplus 2)_{\rho\pi}\oplus(1 \oplus 2\oplus 3)_{(\pi\pi)\pi}\,,
\end{equation}
where the numbers on the right-hand side indicate total three-pion isospins $I_{\pi \pi \pi}$ and
the subscripts denotes the isospins of the two-particle subsystem, $I_{\pi \pi}$. In particular $\sigma$ refers to $I_{\pi \pi} = 0$, $\rho$ to $I_{\pi \pi} = 1$ and $(\pi\pi)$ to $I_{\pi \pi} = 2$.
Projecting to definite isospin results in four independent quantization conditions.

\item[(iv)] Truncation to a maximum angular momentum $\ell_{\rm max}$. Due to the mixing of partial waves (a consequence of the reduced symmetry of the finite volume), the quantization condition formally depends on an infinite number of partial waves. To make the problem tractable, the matrices $\mathcal K_2$ and $\mathcal{K}_{{\rm df},3}$ must be truncated, such that their entries vanish whenever $\ell >\ell_{\rm max}$. This work focuses on lowest nonzero two-particle angular momenta: $\ell_{\rm max} = 1$ for $I_{\pi \pi} = 1$ and $\ell_{\rm max} = 0$ for $I_{\pi \pi} = 0,2$.

\item[(v)] Parametrization the two- and three-particle K-matrices, in order to express the quantization condition completely as a function of $E, \boldsymbol P, L$, together with a set of K-matrix parameters. Numerically determining the roots in $E$ with all other parameters fixed provides the energy spectrum for the three-pion system under the assumption of this model. In this work we restrict attention to the case of vanishing three-particle interactions, $\mathcal{K}_{{\rm df},3} = 0$. The two-particle interactions are given by
\begin{gather}
\frac{p}{m_\pi} \cot{\delta}_{\sigma}(p) = \frac{6\pi}{g_\sigma^2} \frac{m_\sigma^2 - E^2}{Em_\pi} \frac{E^2}{m_\sigma^2} \,, \qquad \frac{p^3}{m^3_\pi} \cot{\delta}_\rho(p) = \frac{6\pi}{g_\rho^2} \frac{m_\rho^2 - E^2}{Em_\pi} \frac{E^2}{m_\rho^2} \,, \\ \frac{p}{m_\pi} \cot{\delta}_{\pi \pi}(p) = - \frac{1}{m_\pi a_{\pi \pi}} \,.
\end{gather}
Thus, our energies depend on up to 5 parameters: $m_\sigma, m_\rho, g_\sigma,g_\rho$ and $a_{\pi \pi}$. Here we do not count $m_\pi$ separately as all quantities are presented in units of $m_\pi$.

We stress that our aim here is not to provide realistic parametrizations of the interactions, but to illustrate how, for a given set of interactions, the finite-volume energies can be predicted. In a practical lattice QCD calculation, $\mathcal K_2$ will be determined from a combined fit to lattice-QCD determined, two- and three-pion energies.
\end{itemize}

\section{Finite volume energies}

In this talk, we highlight a subset of a larger set of results, which will be presented more completely in a subsequent publication.

\begin{figure}[t]
\centering
\begin{minipage}{0.32\textwidth}
\includegraphics[width=\linewidth]{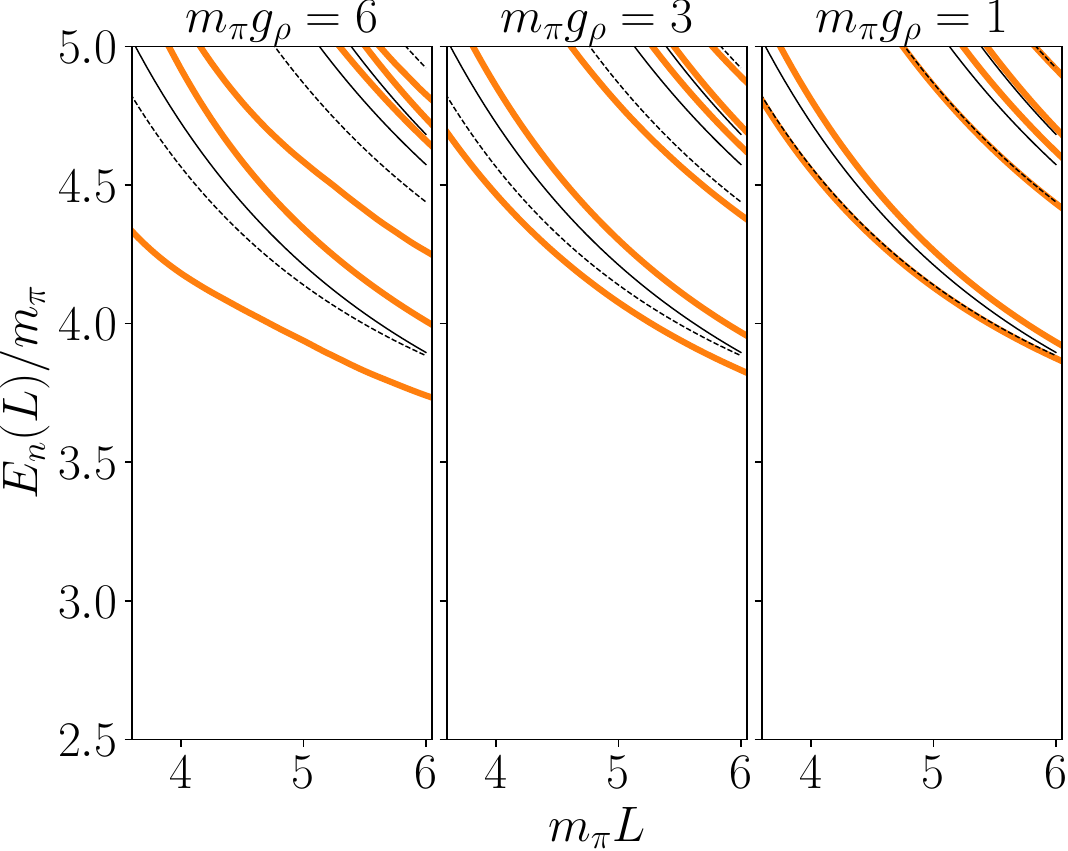}
\subcaption{}
\end{minipage}
\hfill
\begin{minipage}{0.32\textwidth}
\includegraphics[width=\linewidth]{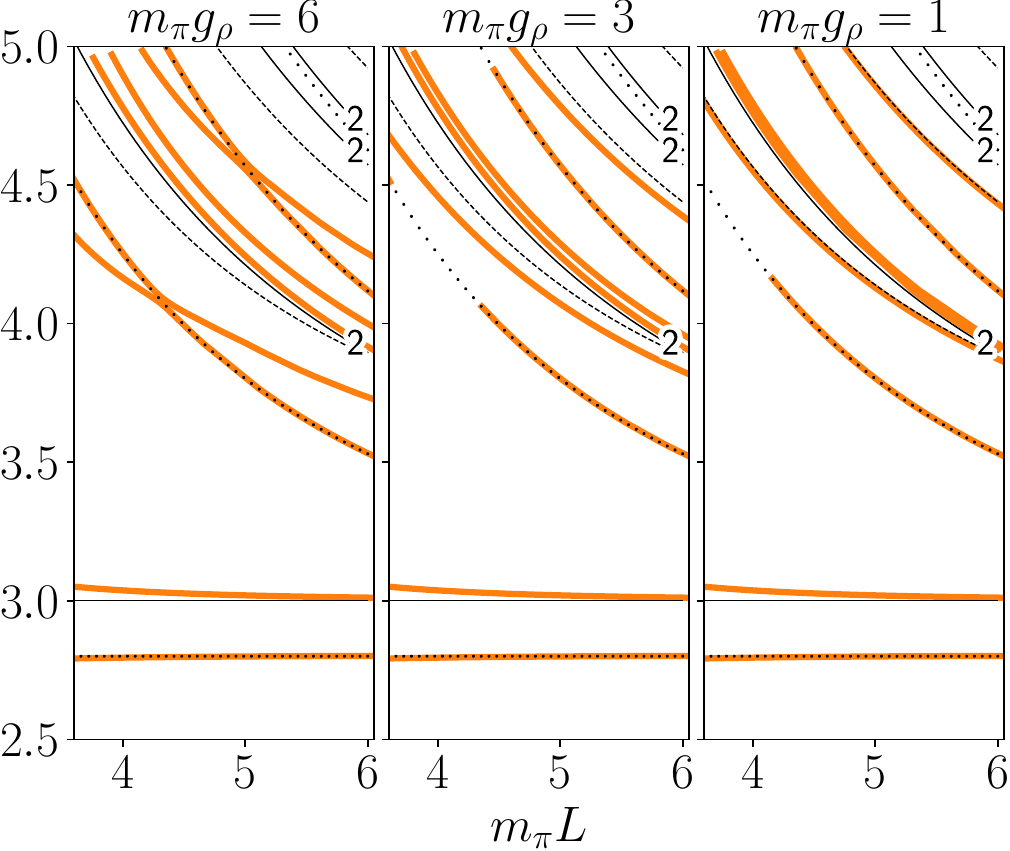}
\subcaption{}
\label{fig:iso1}
\end{minipage}
\hfill
\begin{minipage}{0.32\textwidth}
\includegraphics[width=\linewidth]{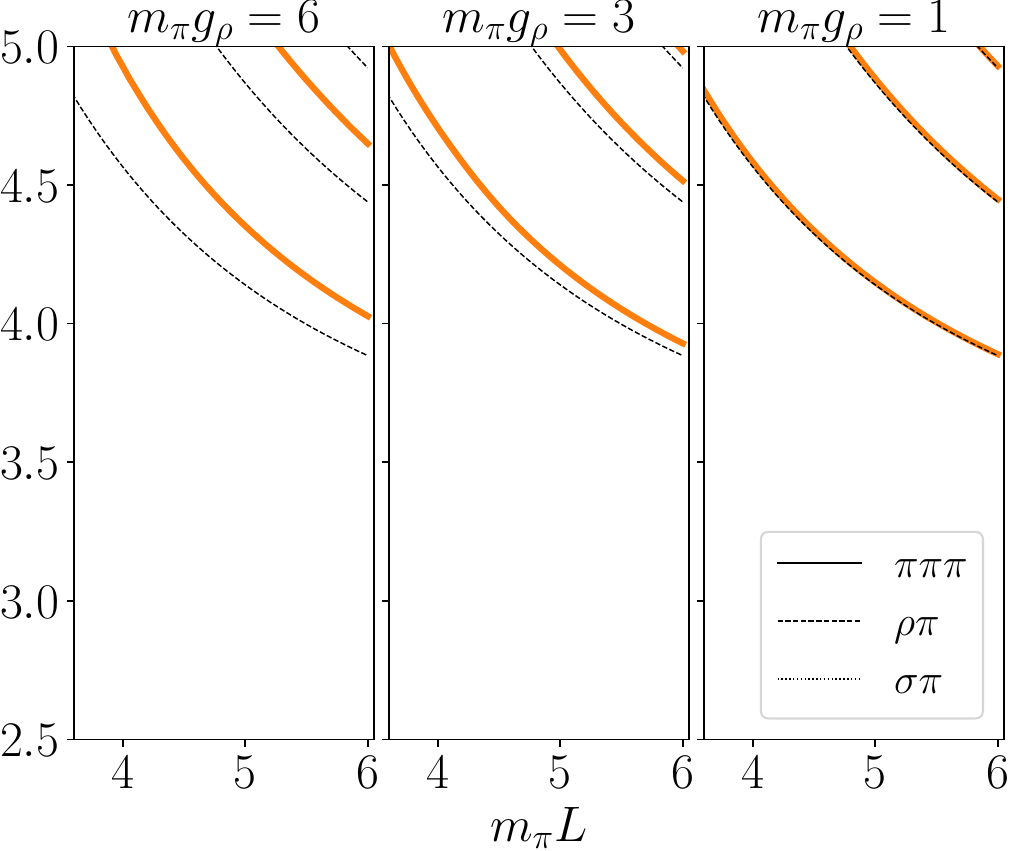}
\subcaption{}
\end{minipage}
\caption{The interacting finite volume energies \textit{(solid orange lines)} of irrep $A_1^-$ in isospin channel: (a) $I_{\pi\pi\pi}=2$, (b) $I_{\pi\pi\pi}=1$ and (c) $I_{\pi\pi\pi}=0$. The non-interacting energies are illustrated in black \textit{solid}, \textit{dashed} and \textit{dotted} lines for $\pi\pi\pi$, $\rho\pi$ and $\sigma\pi$ states, respectively. The small black numbers give the multiplicity of non-interacting states in the cases where this is greater than one. The three panels in each subplot correspond to increasing values of the $\rho$ coupling. All other parameters are held fixed, in particular with $m_\rho/m_\pi = 2.2$, $m_\pi g_\sigma = 1.0$, $m_\sigma/m_\pi = 1.8$ and $m_\pi a_{\pi \pi} = 0.1$. We do not claim that these parameters (nor the underlying parametrizations) illustrate a realistic lattice QCD system, but only serve to illustrate the method.\label{fig:A1MINUS_irrep}}
\end{figure}

We begin with figure~\ref{fig:A1MINUS_irrep}, which shows the finite-volume energies in the $A_1^-$ irrep. The figure illustrates how the interacting spectrum is shifted from the non-interacting energies of $\pi\pi\pi$, $\rho\pi$, and $\sigma\pi$ states. In each panel only the value of the $\rho$ coupling is varied with all other parameters held fixed. The levels exhibit various interesting phenomena, such as avoided level crossings between $\rho \pi$ like levels and $\sigma \pi$ like levels in figure~\ref{fig:A1MINUS_irrep}(b). In a lattice QCD calculation, energies in the vicinity of such an avoided level crossing are expected to be particularly sensitive to the $\rho \pi \to \sigma \pi$ interaction. We emphasize here that the resonance nature, both of the $\sigma$ and of the $\rho$ is being rigorously incorporated through the RFT formalism.

\begin{figure}[h]
\centering
\begin{minipage}{0.33\textwidth}
\includegraphics[width=\linewidth]{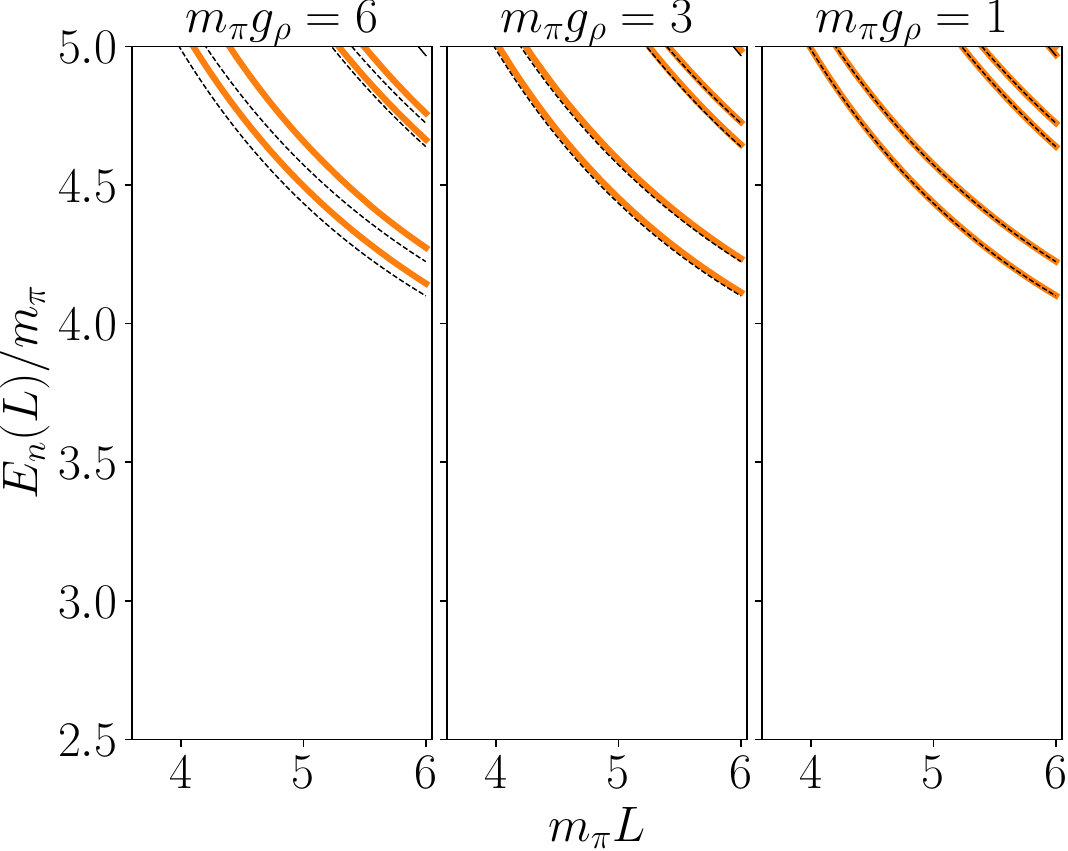}
\subcaption{}
\end{minipage}
\hspace{1cm}
\begin{minipage}{0.33\textwidth}
\includegraphics[width=\linewidth]{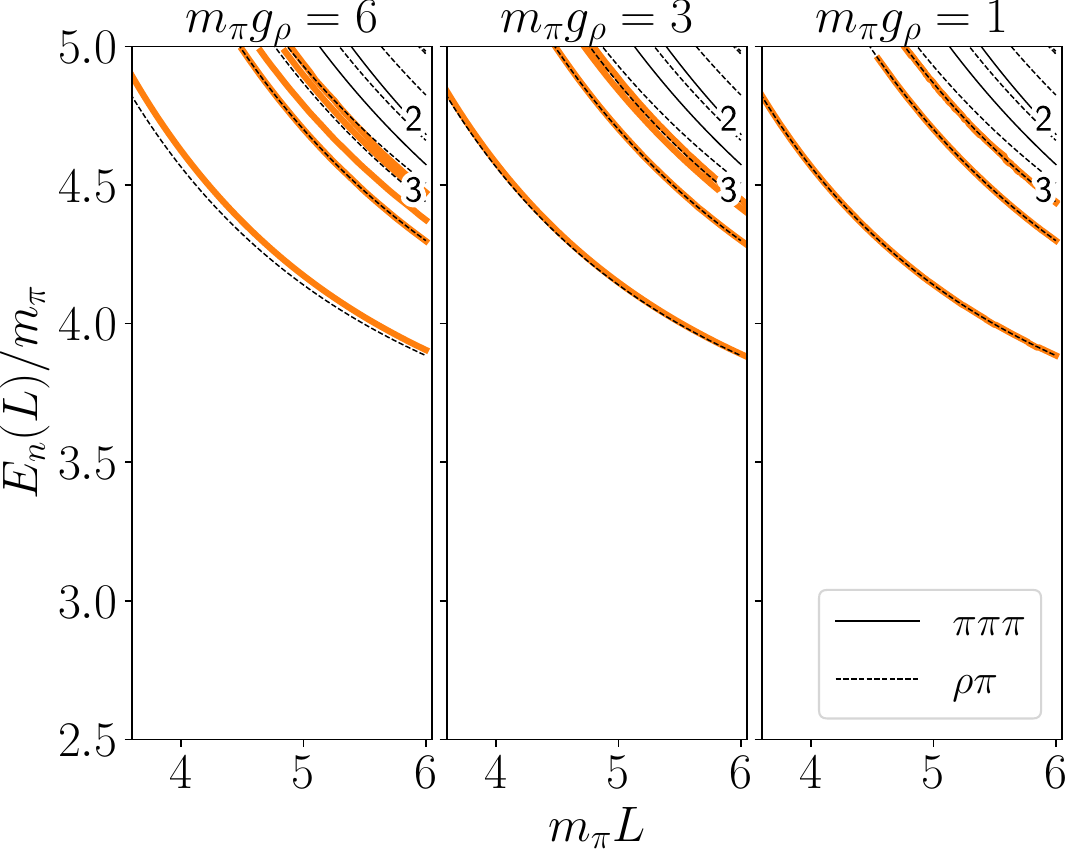}
\subcaption{}
\end{minipage}
\caption{The $A_1$ energy spectrum in $I_{\pi\pi\pi} = 2$ for moving frames with total momentum; (a) $\boldsymbol{P}= [001]$, and (b) $\boldsymbol{P}= [011]$. Other features of the plots and other parameter choices are as in figure~\ref{fig:A1MINUS_irrep}.}
\label{fig:A1_irrep}
\end{figure}

Figure \ref{fig:A1_irrep} gives an example of the system at non-zero total momentum. This shows how the states are condensed by adding more momentum to the system. The mixing of partial waves increases with reduced symmetry of the volume. As with the rest frame case, the spectrum shows how the shift from the non-interacting energy grows with the $\rho$ coupling.

\section{Unphysical solutions}

Another result of our numerical investigations is the appearance of unphysical solutions in certain cases. These are manifestly unphysical because they exist only over a finite range of volumes and then disappear as $L$ increases, typically when two curves meet and annihilate. Such behavior has been observed in previous work \cite{\UnphysicalStates}. In certain cases the unphysical states may be due to volume effects that are neglected in the derivation, which decrease faster than any power of $1/L$, but can still be significant depending on the details of the kinematics and interactions.

Another possibility in the present case is that the Breit-Wigner form of $\mathcal K_2$ leads to a subthreshold pole that is not present in the physical scattering amplitude. This is relevant since the quantization condition depends on $\mathcal K_2$ for CMF energies below $2 m_\pi$ down to $0$. The details of the subthreshold dependence are determined by a cutoff function that varies between $0$ and $1$ in the region $0 < (E^\star_{2,k})^2 < (2m_\pi)^2$.

\begin{figure}[h]
\begin{minipage}{.3\textwidth}
\hfill
\end{minipage}
\begin{minipage}{.7\textwidth}
\hspace{-1.3cm}
\centering
\includegraphics[scale=0.49]{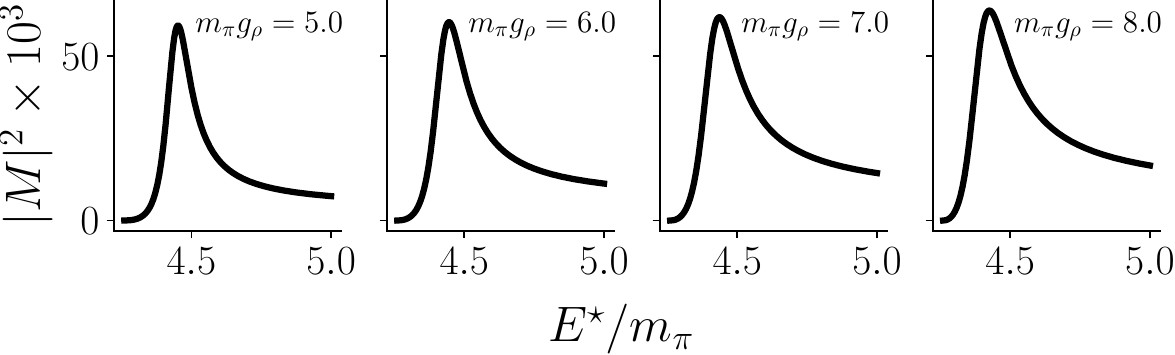}
\end{minipage}
\begin{minipage}{.3\textwidth}
\centering
\includegraphics[scale=0.5]{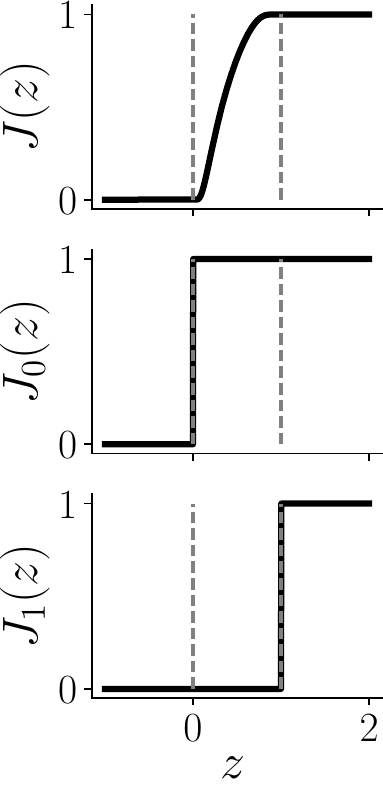}
\end{minipage}%
\begin{minipage}{.7\textwidth}
\hspace{-1.3cm}
\centering
\includegraphics[scale=0.5]{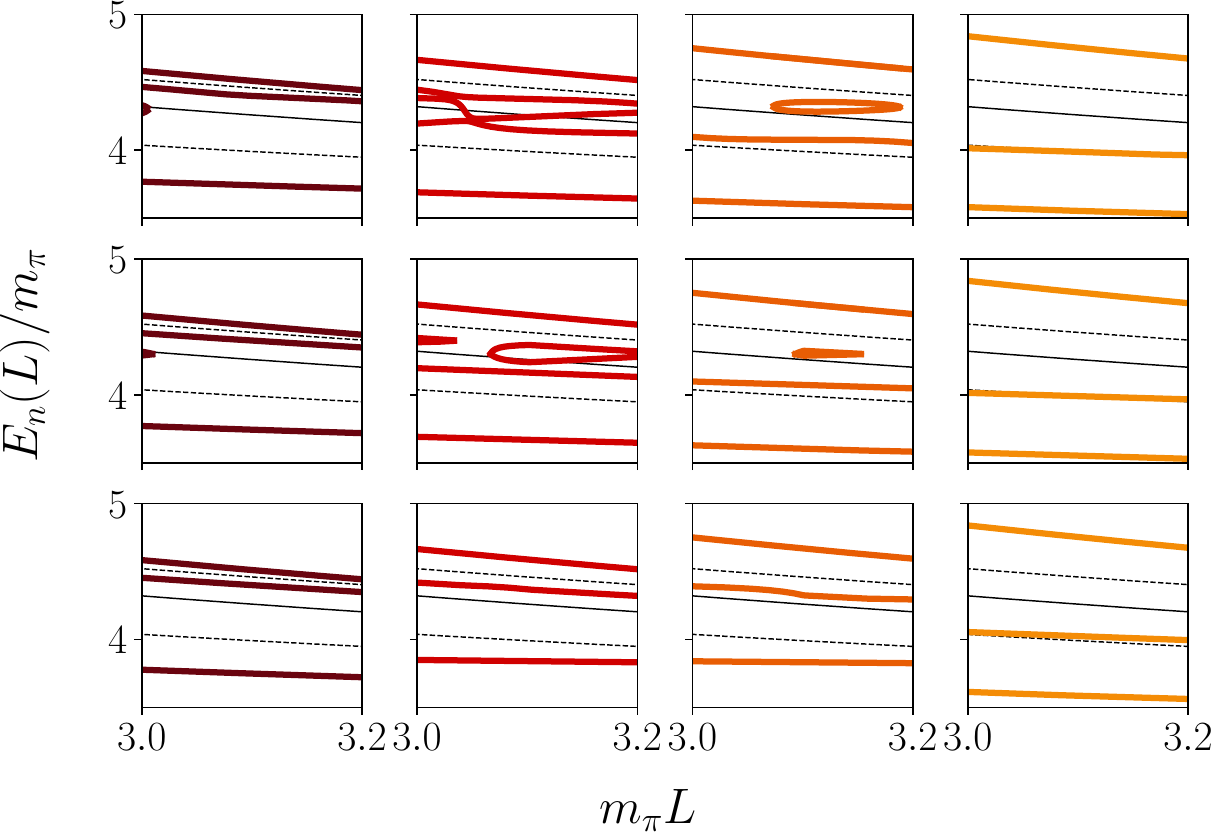}
\end{minipage}
\caption{The finite-volume energies of $A_2$ irrep in isospin channel $I_{\pi\pi\pi} =2$ and $\boldsymbol{P}=(2\pi/L)[001]$ frame. The spectrum is extracted with increasing $g_\rho$ coupling, shown in the top panel, and using three cutoff functions, shown in the left panel.}
\label{fig:unphysical}
\end{figure}

In figure~\ref{fig:unphysical} we show an example of unphysical behavior for $I_{\pi\pi\pi}=2$ with $\boldsymbol P = [001]$ in the $A_2$ irrep. Note that the issue in this case appears for very small volumes and is therefore not particularly concerning. Given the spurious subthreshold poles, we were motivated to investigate how the unphysical behavior depends on the cutoff function (varied over the three rows) as well as the strength of interactions (varied over the four columns).

In all previous examples, our choice of cutoff function follows ref.~\cite{Hansen:2014eka}, giving a smooth transition between $0$ and $1$ as shown in the top right panel, where $z \equiv (E^\star_{2,k})^2/(4m_\pi^2)$. This is the choice made in the top row of figure~\ref{fig:unphysical}. For the bottom two rows we take a different profile of the cutoff as shown. We note that the spurious states are removed in the bottom row, but at the cost of introducing power-like volume effects from the step. Further investigation is required to understand the best solution to this problem in general.

\section{Summary and outlook}

In this talk, we have presented progress in implementing the three-particle RFT formalism to relate K-matrices to discretized finite-volume energies $E_n(\boldsymbol{P},L)$ for all non-maximal-isospin three-pion channels $I_{\pi\pi\pi} = 2,1,0$. The results focus on vanishing three-particle interaction and provide a benchmark for future lattice QCD calculations of three-pion resonances. The energies were numerically evaluated using the open-source Python package {\sf ampyL}~\cite{ampyL}.

A natural extension of this work is the implementation of the quantization condition for non-zero $\mathcal{K}_{{\rm df},3}$. Chiral effective theory can serve as a useful tool to parameterize the interactions in this case, as recently discussed in ref.~\cite{Baeza-Ballesteros:2024mii}. Other generalizations include the extension of our numerical implementation to non-identical and non-degenerate particles, such as systems with pions and kaons, and particles with non-zero intrinsic spin. By implementing the formulas in an efficient and robust open-source library, we aim to pave the way for an efficient workflow that facilitates the extraction of three-particle scattering amplitudes from lattice-QCD-determined finite-volume energies, an approach that has already seen major success in the two-particle sector.

\acknowledgments

We thank
Ra{\'u}l Brice{\~n}o,
Fernando Romero-L\'opez,
Steve Sharpe,
and
Christopher Thomas
for useful discussions. MTH is supported in part by UK STFC grants
ST/P000630/1
and
ST/X000494/1, and additionally by UKRI Future Leader Fellowship MR/T019956/1.
The work of Athari Alotaibi is funded by King Saud University (Riyadh, Saudi Arabia).

\bibliographystyle{JHEP}
\bibliography{refs.bib}

\end{document}